\documentclass[12pt]{article}
\usepackage{latexsym}
\usepackage{epsfig}

\usepackage{graphicx}
\usepackage{epstopdf}

\usepackage{epsfig,amssymb,amsmath,amscd}

\newcommand{\bq}{\begin{equation}}
\newcommand{\eq}{\end{equation}}
\newcommand{\bqa}{\begin{eqnarray}}
\newcommand{\eqa}{\end{eqnarray}}
\newcommand{\ben}{\begin{enumerate}}
\newcommand{\een}{\end{enumerate}}
\newcommand{\bc}{\begin{center}}
\newcommand{\ec}{\end{center}}
\newcommand{\bqb}{\begin{eqnarray*}}
\newcommand{\eqb}{\end{eqnarray*}}

\def\pr#1#2#3{Phys. Rev. ${\bf{#1}}$, #2 (#3)}

\def\pl#1#2#3{Phys. Lett. ${\bf{#1}}$, #2 (#3)}

\def\np#1#2#3{Nucl. Phys. ${\bf{#1}}$, #2 (#3)}

\def\jmp#1#2#3{J. Mod. Phys. ${\bf{#1}}$, #2 (#3)}

\begin{document}
\pagenumbering{arabic}
\thispagestyle{empty}
\def\thefootnote{\fnsymbol{footnote}}
\setcounter{footnote}{1}

\vspace*{2cm}
\begin{flushright}
August 16, 2018\\
%arXiv: \\
 \end{flushright}
\vspace*{1cm}

\begin{center}
{\Large {\bf
Further tests of special interactions of massive particles from the 
Z polarization rate in $e^+e^-\to Zt\bar t$ and in $e^+e^-\to ZW^+W^-$ 
.}}\\
 \vspace{1cm}
%-----------------------------------------------------------------
{\large F.M. Renard}\\
\vspace{0.2cm}
Laboratoire Univers et Particules de Montpellier,
UMR 5299\\
Universit\'{e} de Montpellier, Place Eug\`{e}ne Bataillon CC072\\
 F-34095 Montpellier Cedex 5, France.\\
\end{center}

\vspace*{1.cm}
\begin{center}
{\bf Abstract}
\end{center}

We propose further tests of the occurence of scale dependent heavy 
particle masses (Z,W,t) and of strong final state interactions by comparing 
Z longitudinal polarization rates in different kinematical distributions
of the $e^+e^-\to Zt\bar t$ and in $e^+e^-\to ZW^+W^-$ processes.

\vspace{0.5cm}

\def\thefootnote{\arabic{footnote}}
\setcounter{footnote}{0}
\clearpage

\section{INTRODUCTION}

In previous papers \cite{eettZ,ggttZ,Wtb,eeZWW} we have shown that the
rate of $Z_L$ polarization in several $Zt\bar t$ and $ZWW$ production processes
is directly sensitive to the occurence of scale dependent masses (see \cite{trcomp, CSMrev}) 
and of final state interactions between heavy particles (for example due to a 
substructure \cite{comp, Hcomp2,Hcomp3,Hcomp4,partialcomp} like in the 
hadronic case or to a dark matter environment \cite{revDM,DMmass, DMexch}).\\
We now want to improve these tests by looking at different kinematical distributions
of the $Z_L$ rate and by comparing the effects in the $Zt\bar t$ and $ZWW$ production 
processes in order to identify the origin of the effects, pure $t$, pure $Z,W$ or both.\\
We will concentrate on the $e^+e^-\to ZW^°W^-$ and $e^+e^-\to Zt\bar t$  processes
and illustrate the effects on the distributions of the $Z_L$ rates versus different 
final 2-body invariant energies.\\
The SM properties (from the respective Born diagrams) have been recalled in 
\cite{eettZ,eeZWW} and illustrated for the $p_Z$ distribution.
The sensitivity of the $Z_L$ rate to  the concerned masses is natural in SM
due to the Goldstone equivalence \cite{equiv}.\\

In this paper
we will first compute the corresponding $s_{WW}$, $s_{ZW^+}$, $s_{ZW^-}$
and $s_{t\bar t}$, $s_{Zt}$, $s_{Z\bar t}$ distributions of the $Z_L$ rate.
Like in the previous papers we will then introduce the two different types
of modifications, scale dependent top quark, $Z$, $W$ masses and 2-body possible
final state interactions. Illustrations will be made with simple kinematical
dependences but one can easily imagine what would give more elaborated 
forms like resonances with Breit-Wigner forms.\\
In Section 2 we consider the $e^+e^-\to ZW^°W^-$ process with scale dependent
$Z,W$ masses (keeping $c_W$ at its SM value) and ${WW}$, ${ZW^+}$, ${ZW^-}$
final state interactions.\\
In Section 3 we consider the $e^+e^-\to Zt\bar t-$ process with scale dependent
masses for the $Z$ or the $t$ or both and ${t\bar t}$, ${Zt}$, ${Z\bar t}$ 
final state interactions.\\
In Section 4 we will conclude by summarizing the informations that 
may be obtained from the comparison of the two processes, in particular 
about the simultaneous occurence or not of the scale dependence of the 
top quark mass and of the $Z,W$ masses.\\

\section{$e^+e^-\to ZW^°W^-$}

The Born SM diagrams have been given in \cite{eeZWW} with illustration of
the $p_Z$ distribution for the $Z_L$ rate

\bq
R_L={\sigma(Z_L WW)\over \sigma(Z_T WW)+\sigma(Z_L WW)}~~\label{RLw}
\eq

The scale dependence of the $Z,W$ masses has been studied (assuming that the 
$m_W/m_Z$ ratio (i.e. $c_W$) is fixed) with the test form

\bq
m_W(s)=m_W{(m^2_{th}+m^2_0)\over (s+m^2_0)}~~\label{mw}
\eq

Effects of final state interactions were illustrated by multiplying the
amplitudes by the $(1+C(s_{ZW^+})) (1+C(s_{ZW^-})) (1+C(s_{W^+W^-}))$ 
"test factor"
with
\bq
C(x)=1+{m^2_{Z}\over m^2_0}~ln{-x\over (m_Z+m_W)^2} ~~, \label{Cxw}
\eq

In Fig.1 (up) we plot the $s_{WW}$ distribution of the $Z_L$ rate for $\sqrt{s}= 5$ TeV 
and $\theta=\pi/2$.
It is directly related to the $p_Z$ distribution shown in \cite{eeZWW} as
$s_{WW}=s+m^2_Z-2E_Z\sqrt{s}$.\\
In Fig.1 (down) we plot the $s_{ZW^+}$ distribution for the same kinematical conditions;
we do not show the  $s_{ZW^-}$ distribution which is very similar.\\
In both cases we can see the basic SM contributions and the effect of a modification of the
$Z,W$ masses according to eq.(\ref{mw}). The shapes of the distributions and of
their modifications are typically different  in the $s_{WW}$ and in the $s_{ZW^{\pm}}$
cases.\\

In Fig.2 we then show, with the same conditions, the effects of final state interactions
according to eq.(\ref{Cxw}) and as in \cite{eeZWW} from the addition of the $Z$ and of the $G^0$
intermediate contributions. We can also see the differences between the shapes of these distributions
and between the ones due to scale dependent masses or final interactions.\\
With other types of "test forms" the differences could  even be
stronger and specific of the origin of these new interactions (for example with resonance
contributions).\\

\section{$e^+e^-\to Zt\bar t$}

The behaviour of the $Z_L$ rate
\bq
R_L={\sigma(Z_L t\bar t)\over \sigma(Z_T  t\bar t)+\sigma(Z_L  t\bar t)}~~\label{RLt}
\eq
in this process has been studied in 
\cite{eettZ} where one can find the SM diagrams and the corresponding $p_Z$ distributions.

In addition to the scale dependence of the $Z,W$ masses one may now have a scale 
dependence of the top quark mass that we will similarly study with the test
form

\bq
m_t(s)=m_t{(m^2_{th}+m^2_0)\over (s+m^2_0)}~~\label{mt}
\eq

Final state interactions may now appear differently  between $(Zt)$ or $(Z\bar t)$ and
$(t\bar t)$. So we will separately study their effects with the test factors
affecting the amplitudes respectively:\\
$(1+C(s_{Zt}))$,~~~~ $(1+C(s_{Z\bar t}))$,~~~~and~~ $(1+C(s_{t\bar t}))$\\ 
with

\bq
C(x)=1+{m^2_{t}\over m^2_0}~ln{-x\over (m_Z+m_t)^2} ~~, \label{Cxt}
\eq

Results of scale dependent masses and of final state interactions 
are respectively illustrated in Fig.3 and 4.\\

As expected from the expression of the $Z_L$ polarization vector, a decrease of the
$Z$ mass leads to an increase of the corresponding amplitudes. On another hand
a decrease of the top quark mass leads to a decrease of the longitudinal 
amplitudes; this is expected, by Goldstone equivalence (\cite{equiv}), from the couplings of
the Goldstone boson to the top quark which is proportional to the top quark
mass.\\
Consequently the presence of both $Z$ and $t$ scale dependent masses may cancel and lead 
to almost no visible effect if the forms of the dependences are similar.
This is illustrated in Fig.3 for both $s_{t\bar t}$ and $s_{Zt}$ 
(and similarly $s_{Z\bar t}$).
This is the remarkable feature of this process.\\
For comparison we then show, in Fig.4, the effects of specific final state
interactions on the $s_{t\bar t}$ and $s_{Zt}$ distributions.
We separately illustrate the effects of $s_{t\bar t}$ interactions (label $t$),
of $s_{Zt}$ and $s_{Z\bar t}$ interactions (label $Z$), and of all of them
(label $Zt$) giving progressively stronger effects and again specific shapes
as compared to the above ones.\\

\section{Conclusion}

In this paper we have made a comparative study of the longitudinal $Z$ polarization
rate in the $e^+e^-\to Zt\bar t$ and $e^+e^-\to ZW^°W^-$ processes; this has shown 
its remarkable richness.\\
In $ZWW$ production this rate is directly controlled by the $W$ and $Z$ masses;
the  $W$ mass dependence occurs in the $ZGW$ couplings and both the $W$ and $Z$ masses in the
respective polarization vector. We assumed that the SM structure is
maintained ($m_W/m_Z=c_W$) even with scale dependent masses. This leads to an increase
of the $Z_L$ rate as shown in Fig.1.\\
In $Zt\bar t$ production the rate is controlled by both $Z$ and $t$ masses.  
Contrarily to the $ZWW$ case there is no obvious relation between them in SM. 
The $Z$ mass controls the Z polarization vector and the $t$ mass the $Gtt$ couplings
(with $Z_L-G$ equivalence). Their effects are opposite and almost cancel in 
the total $Z_L$ rate (Fig.3).\\
In addition we have shown that the shapes of the $s_{WW}$, $s_{ZW^{\pm}}$,
$s_{t\bar t}$ and $s_{Zt,Z\bar t}$ are kinematically different and differently
affected by masses and by specific final state interactions (Fig.2,4).\\
The illustrations were made with arbitrary choices of parameters controlling the scale 
dependence of the masses and the sizes and energy dependences of the 
final state interactions.
Our figures just show that one may indeed suspect the presence of BSM effects
and guess their type from the behaviours of the $Z_L$ rates, for example those originating
from substructures or from special interactions with a dark matter environment.\\
For experimental possibilities relative to these processes see \cite{ee}.\\
As already mentioned in \cite{eeZWW} other production processes may be interesting 
for confirming possible indications coming from the present proposal,
for example $\gamma-\gamma$, see \cite{gammagamma},
or gluon-gluon in hadronic collisions; for LHC possibilities see 
\cite{lhcContino,lhcRichard}.\\

\newpage

\begin{figure}[p]
\vspace{-0cm}
\[
\hspace{-2cm}\epsfig{file=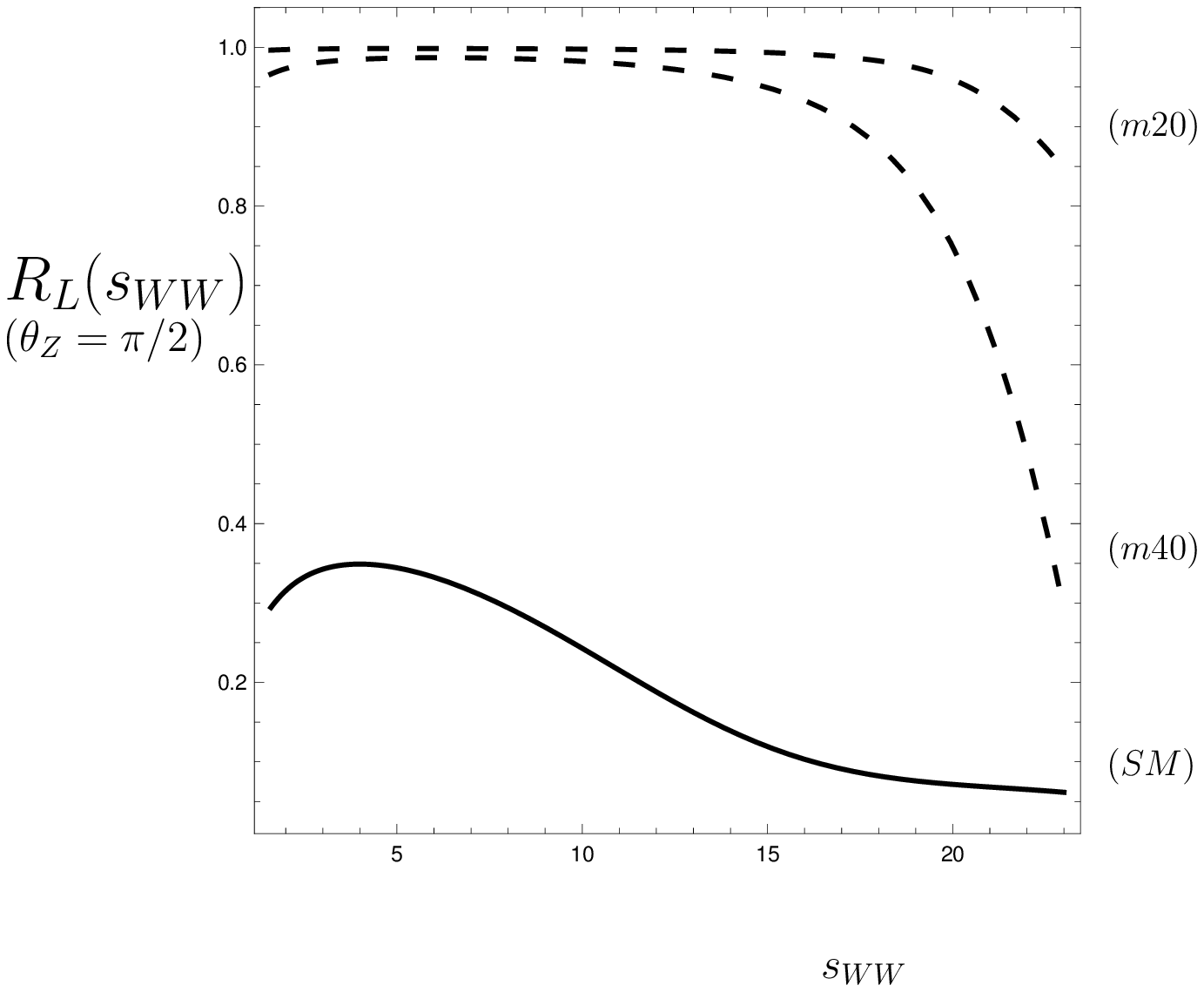 , height=20.cm}
\]\\
\vspace{0.5cm}
\vspace{-13cm}
\[
\hspace{-2cm}\epsfig{file=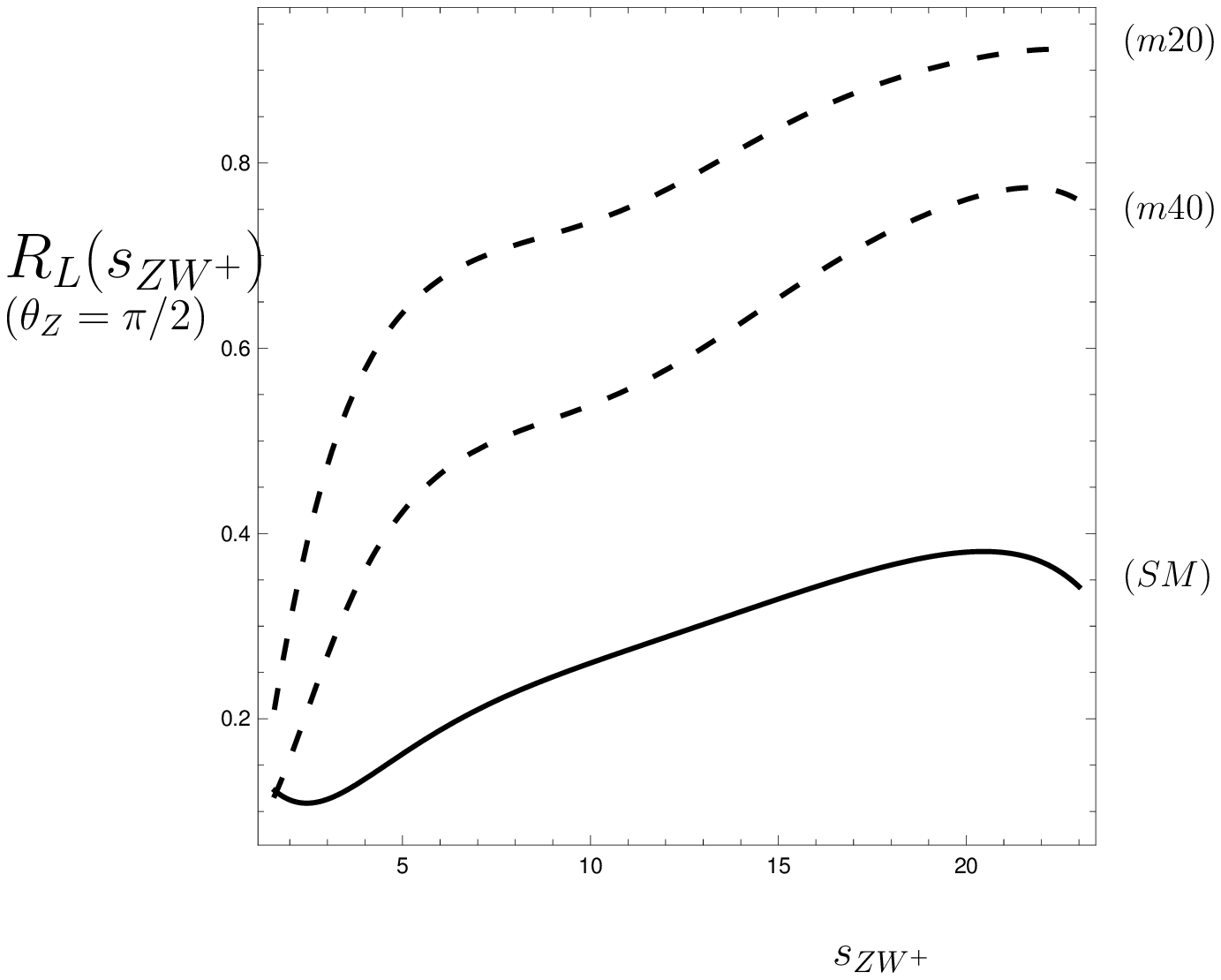 , height=20.cm}
\]\\
\vspace{-10cm}
\caption[1] {$e^+e^- \to Z_LWW $ ratio in SM and 
with an effective $Z$ mass with parameter $m^2_0=20$ or $40$ in eq.(\ref{mw});
invariant distributions for $s_{WW}$ (up) and $s_{ZW}$ (down).}

\end{figure}

\clearpage
\begin{figure}[p]
\vspace{-0cm}
\[
\hspace{-2cm}\epsfig{file=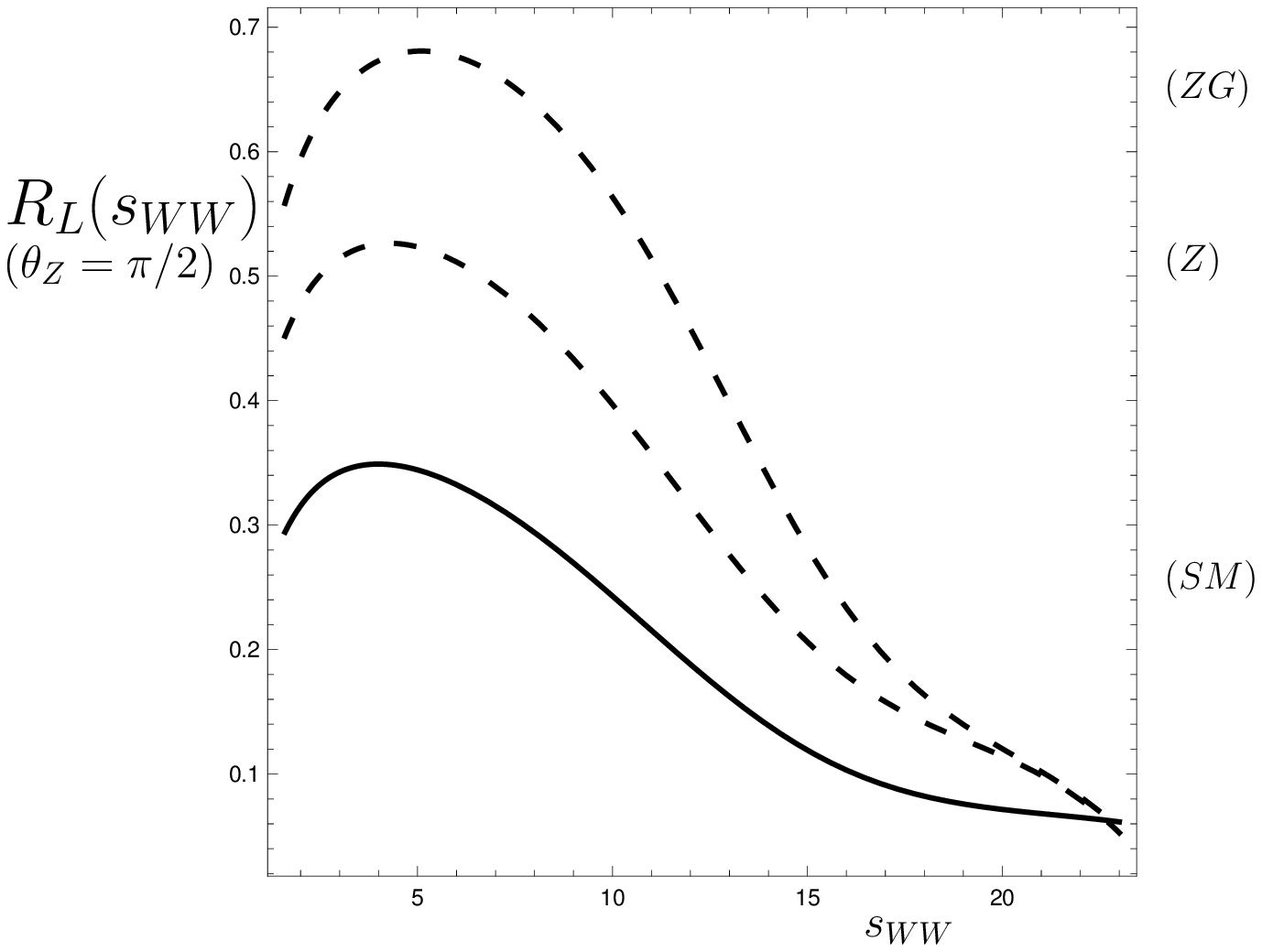  , height=20.cm}
\]\\
\vspace{0.5cm}
\vspace{-13cm}
\[
\hspace{-2cm}\epsfig{file=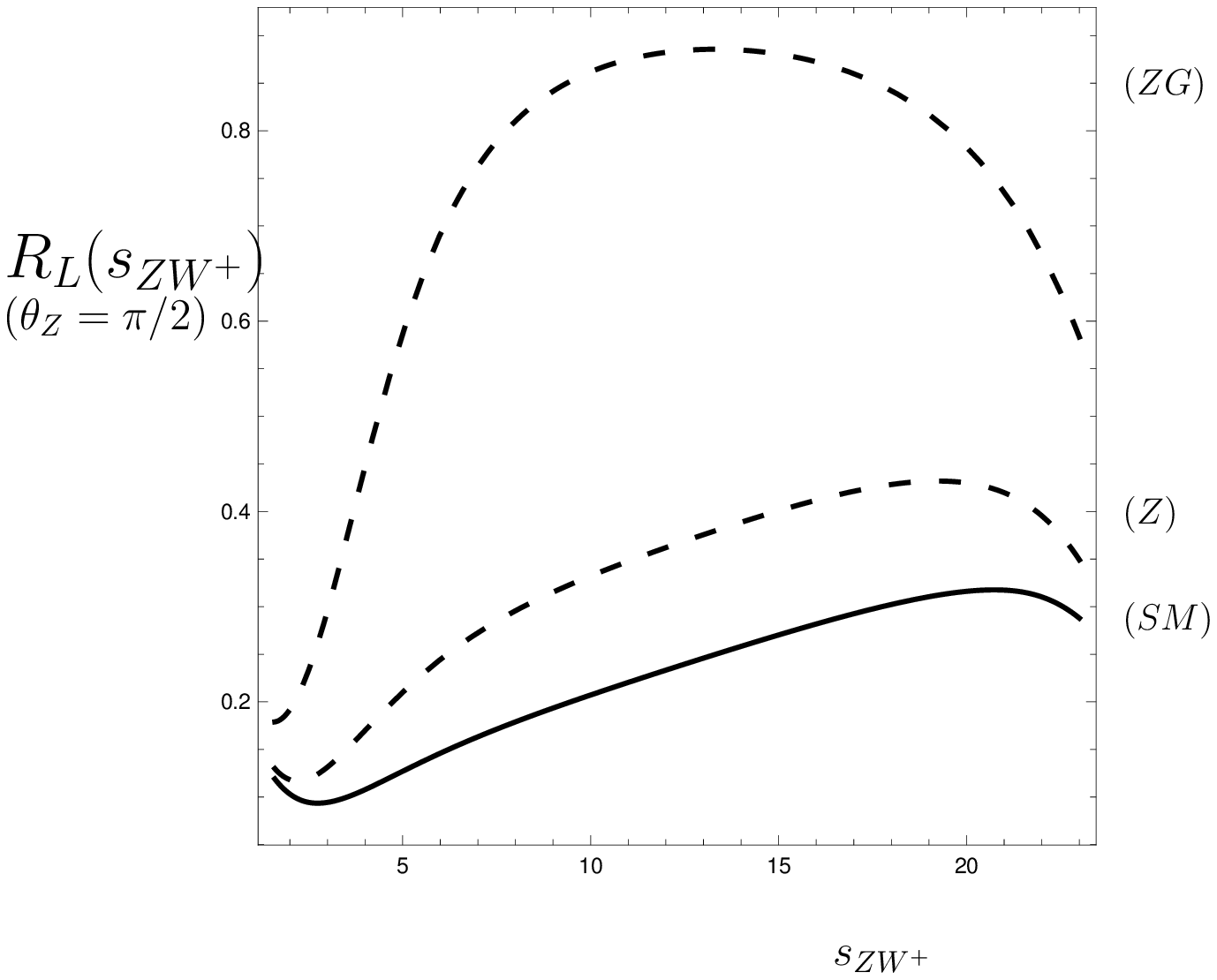  , height=20.cm}
\]\\
\vspace{-10cm}
\caption[1]  {$e^+e^- \to Z_L WW $ ratio in SM and in the cases
of an effective final ($WW$ and $ZW$) interaction (Z) and of an
additional Goldstone contribution ($ZG$) contribution.}

\end{figure}

\begin{figure}[p]
\vspace{-0cm}
\[
\hspace{-2cm}\epsfig{file=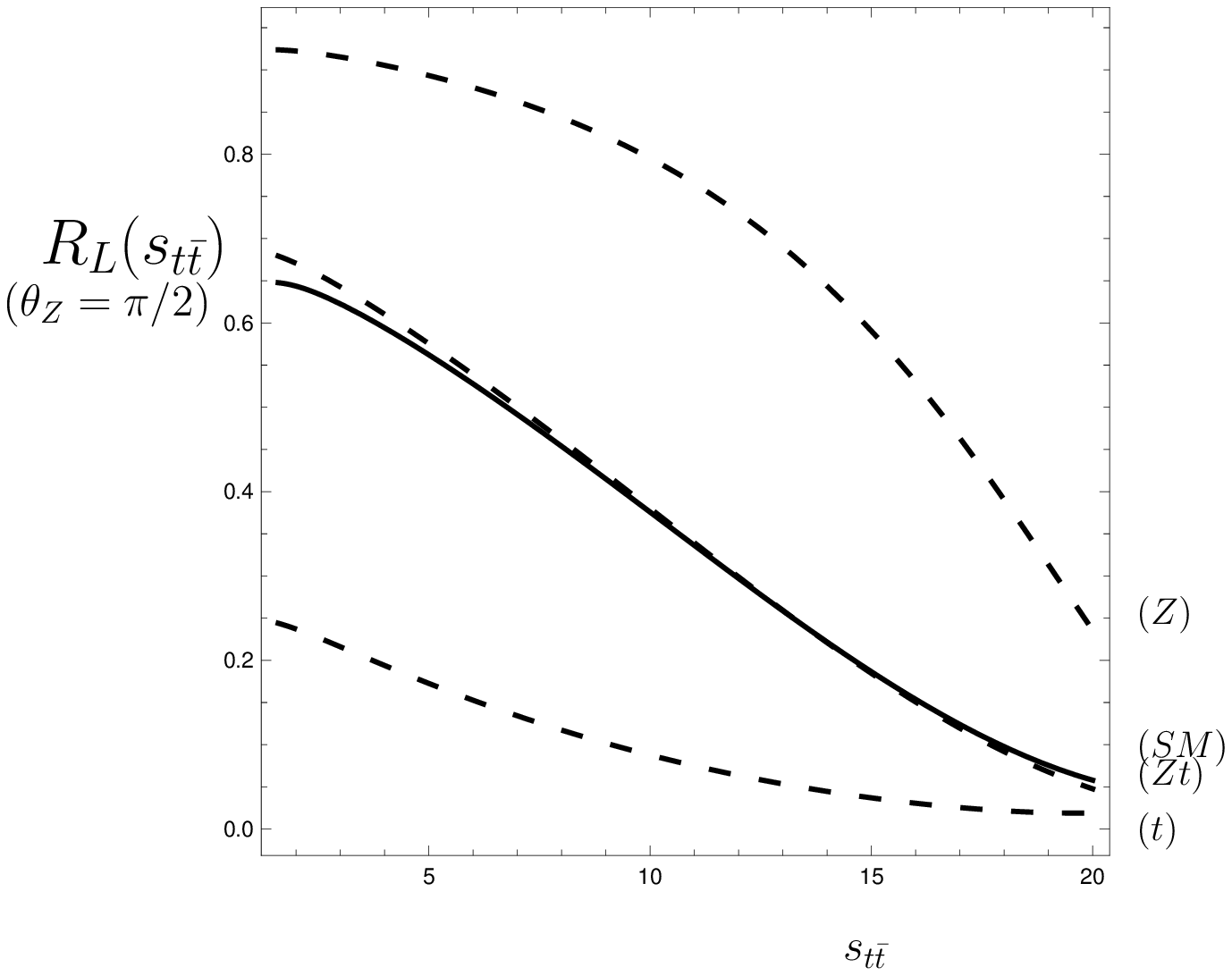 , height=20.cm}
\]\\
\vspace{0.5cm}
\vspace{-13cm}
\[
\hspace{-2cm}\epsfig{file=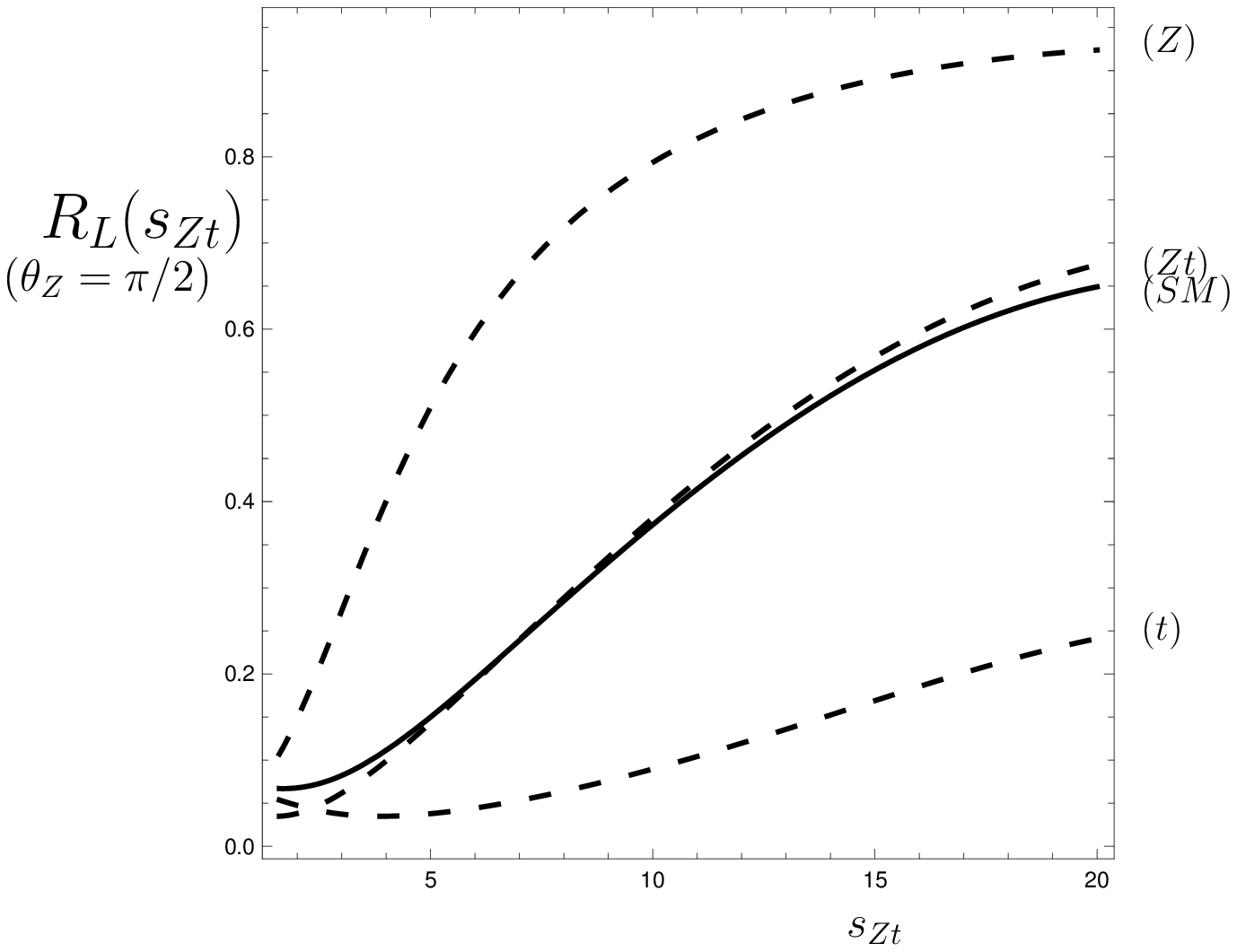 , height=20.cm}
\]\\
\vspace{-10cm}
\caption[1] {$e^+e^- \to Z_L t\bar t $ ratio in SM and in the cases
of an effective top mass (t), of an effective $Z$ mass (Z) and of both (Zt);
invariant distributions for $s_{t\bar t}$ (up) and $s_{Zt}$ (down).}

\end{figure}

\clearpage
\begin{figure}[p]
\vspace{-0cm}
\[
\hspace{-2cm}\epsfig{file=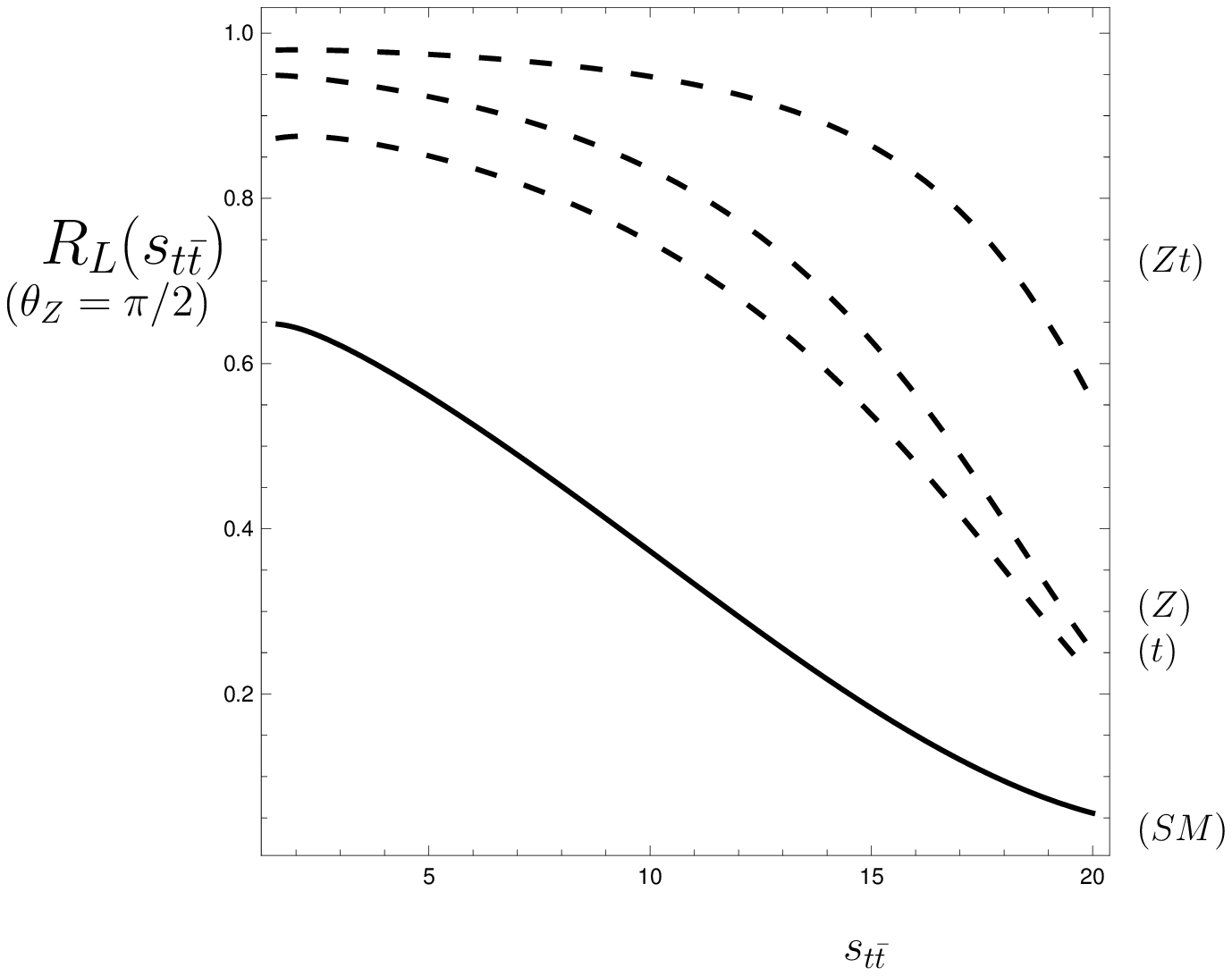 , height=20.cm}
\]\\
\vspace{0.5cm}
\vspace{-13cm}
\[
\hspace{-2cm}\epsfig{file=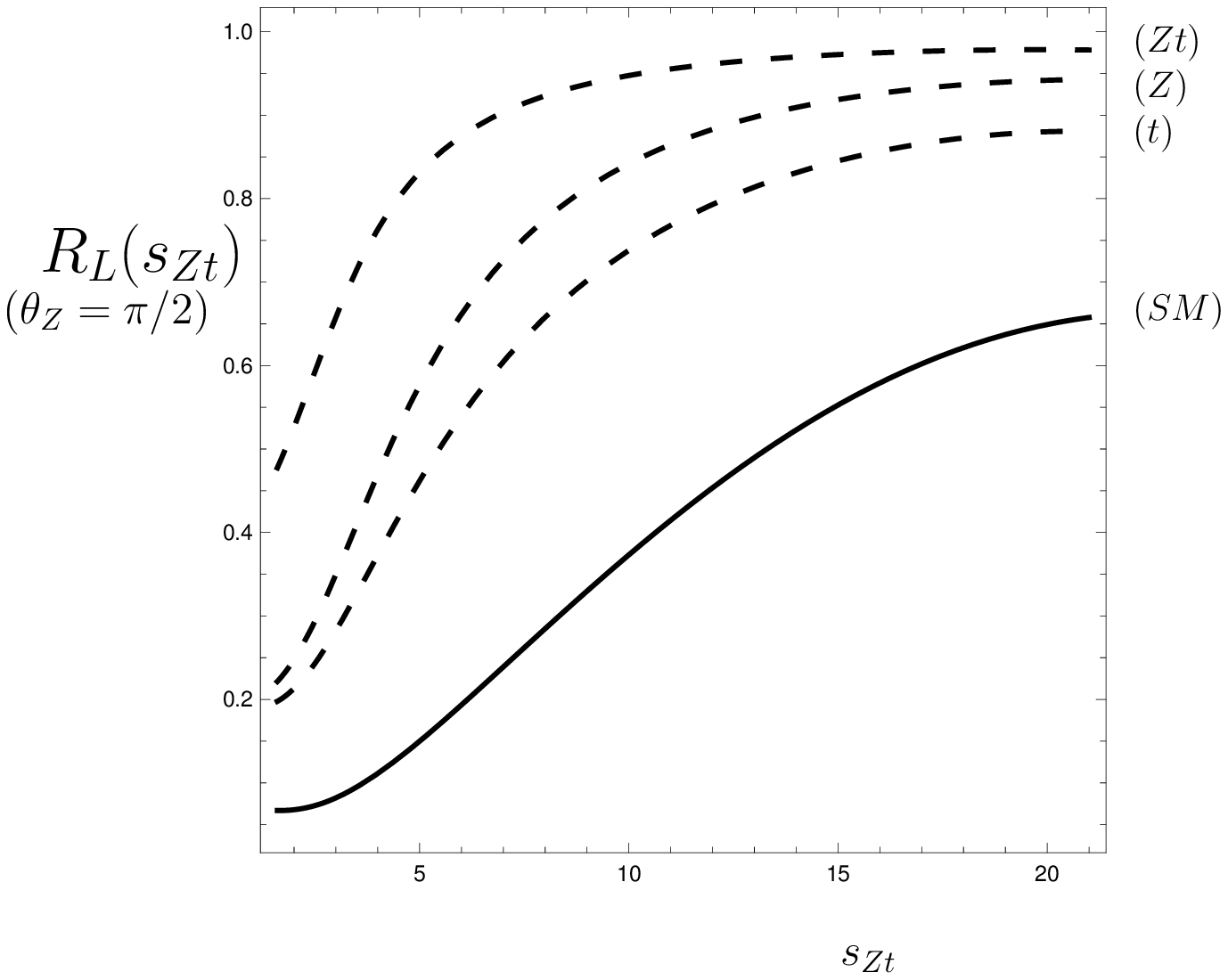 , height=20.cm}
\]\\
\vspace{-10cm}
\caption[1]  {$e^+e^- \to Z_L t\bar t $ ratio in SM and in the cases
of an effective final ($t\bar t$) interaction (t), of an effective final ($Zt$
and $Z\bar t$) interaction ($Z$) and of both ($Zt$).}
\end{figure}

\end{document}